\documentclass[aps,pra,twocolumn,superscriptaddress,longbibliography]{revtex4-1}

\usepackage{natbib}
\usepackage{pgfplots}
\usepackage[breaklinks]{hyperref}
\usepackage{amssymb,bbold,bm}
\usepackage{amsmath,amsthm}
\usepackage{amsfonts,dsfont}
\usepackage{graphicx}
\usepackage{mathtools}
\usepackage{tikz}
\usepackage[english]{babel}
\usetikzlibrary{positioning}

\usepackage[normalem]{ulem}
\usepackage{color}
\usepackage{xcolor}

\newcommand{\g}{\gamma}

\newcommand{\G}{\Gamma}

\newcommand{\F}{\mathcal{F}}

\newcommand{\Tr}{\mbox{Tr}}

\newcommand{\bra}[1]{\mbox{$\langle #1 |$}}
\newcommand{\ket}[1]{\mbox{$| #1 \rangle$}}

\begin{document}
\title{Machine Learning Universal Bosonic Functionals}

\author{Jonathan Schmidt}
\affiliation{Institut f\"ur Physik, Martin-Luther-Universit\"at
Halle-Wittenberg, 06120 Halle (Saale), Germany}

\author{Matteo Fadel}
\affiliation{Department of Physics, University of Basel, Klingelbergstrasse 82, 4056 Basel, Switzerland}

\author{Carlos L. Benavides-Riveros}
\email{carlosbe@pks.mpg.de}
\affiliation{Max Planck Institute for the Physics of Complex Systems, N\"othnitzer Str.~38, 01187, Dresden, Germany}
\affiliation{NR-ISM, Division of Ultrafast Processes in Materials (FLASHit), Area della Ricerca di Roma 1, Via Salaria Km 29.3, I-00016 Monterotondo Scalo, Italy}

\date{\today}

\begin{abstract}
The one-body reduced density matrix $\gamma$ plays a fundamental role in describing and predicting quantum features of bosonic systems, such as Bose-Einstein condensation. The recently proposed reduced density matrix functional theory for bosonic ground states establishes the existence of a universal functional $\mathcal{F}[\gamma]$ that recovers quantum correlations exactly. Based on a novel decomposition of $\gamma$, we have developed a method to design reliable approximations for such universal functionals: our results suggest that for translational invariant systems  the constrained search approach of functional theories can be transformed into an unconstrained problem through a parametrization of an Euclidian space. This simplification of the search approach allows us to use standard machine-learning methods to perform a quite efficient computation of both $\mathcal{F}[\gamma]$ and its functional derivative. For the Bose-Hubbard model, we present a comparison between our approach and Quantum Monte Carlo. 
\end{abstract}

\maketitle

In 1964 Hohenberg and Kohn proved the existence of a universal functional $\mathcal{F}[\rho]$ of the particle density $\rho$, that captures the exact electronic contribution to the ground-state energy of a system of interacting electrons \cite{HK}. Due to a remarkable balance of accuracy and computational cost,  first principle modeling of e\-lec\-tro\-nic systems based on the respective Density Functional Theory (DFT) is nowadays a well established daily practice, with great impact in material science, quan\-tum chemistry or condensed matter \cite{Jones15}. For bosonic systems, however, a fully first-principle description has been elusive. This is due in part to the unsuitability of the particle density to describe fundamental bosonic features as orbital occupations, mode entanglement, or non-diagonal order, which are important for predicting and describing bosonic condensation. As a result, the theoretical treatment of inter\-acting bosonic systems mainly relies on exact diagonali\-za\-tion techniques, which are restricted to few tens of or\-bi\-tals \cite{Cazalilla,Chatterjee2020,Lewenstein}, or mean-field theories that are particularly suitable for dilute ultracold gases \cite{PhysRevB.40.546, pita, Dalfovo1999}. Quantum Monte Car\-lo (QMC) is known to be a powerful family of techniques for computing ground-state properties but is still  restricted  due  to  the  fermion sign  problem. While bosonic systems do not suffer such a sign problem, QMC cannot be applied to, e.g., frus\-tra\-ted quantum spin systems \cite{gubernatis_kawashima_werner_2016}. Nowadays, a renewed interest in the ab-initio description of many-body systems has been motivated by the successful application of artificial neural networks to both fermionic and bosonic problems \cite{doi:10.7566/JPSJ.86.093001,PhysRevLett.121.167204}. 

Since the parameters of correlated bosonic systems (ultracold gases, in particular) can be tuned with a high degree  of control, they are  powerful platforms to study  a wide  range  of  model  Hamiltonians,  ranging  from  Hubbard  models  to  bosonic  antiferromagnets \cite{Bloch2008,Chin2010}. They have also become an active research field in the context of quantum simulations \cite{Bloch2005,Gross995,Schafer2020}, and even quantum foundations \cite{Schmied441,Fadel409,Kunkel413,Lange416}. Such need to describe quantum correlations of bosonic systems efficiently has motivated quite recently to put forward a novel physical theory for interacting bosonic systems \cite{BR2020,liebert2020functional}. Based on a generalization of the Hohenberg-Kohn theorem \cite{Gilbert,Pernal2016}, this reduced density matrix functional theory (RDMFT) for bosons establishes the existence of a universal functional $\mathcal{F}_W[\gamma]$ of the one-body reduced density matrix (1RDM): $\g \equiv  N \mbox{Tr}_{N-1}[\G]$, obtained from the $N$-boson density operator $\G$ by integrating out all except one boson, and the two-particle interaction $\hat W$. Since  the 1RDM is  the  natural  variable of the theory, RDMFT is particularly  well-suited for the accurate description of Bose-Einstein condensates (BEC), strongly correlated bosonic systems, or fragmented BEC \cite{PhysRevA.78.023615,PhysRevLett.119.052501}. Furthermore, the information contained in the spectra of the 1RDM can also be sufficient to investigate mul\-ti\-par\-tite quantum correlations in those systems \cite{10.1088/1367-2630/abe15e, Walter1205, Sawicki2014,PhysRevA.86.040304,Yu2021}.
 
Although RDMFT holds the promise of abandoning the complex $N$-particle wave function as the central object, it does not trivialize  the  ground-state  problem.  In fact, the  fundamental  challenge is to provide reliable approximations to the universal interaction functional $\mathcal{F}_W[\gamma]$. Yet, while the Hohenberg-Kohn-type fundational theorem of RDMFT shows the existence of a universal functional, it does not give any indication of  its concrete form. For DFT, the solution to this problem is given in the form of large classes of approximate functionals,  hierarchically organized in the so-called Jacob's ladder. In recent years, the number of such approximation has significantly increased thanks to machine learning \cite{Kalita2021,Brockherde2017,Schmidt2019,PhysRevLett.126.036401,Margraf2021,PhysRevLett.125.076402} and reduced density matrices \cite{Gibney2021} approaches. 

Our work succeeds in providing a strategy on computing approximations for $\mathcal{F}_W[\gamma]$. In this paper we (i) provide an efficient method to capture the essential features of universal functionals for boson lattices, (ii) show how the constrained search approach associated with it can be simplified in the form of an unconstrained problem, and (iii) implement this approach in a standard machine-learning library to compute $\mathcal{F}_W[\gamma]$, its derivative, and the ground-state energy. We shall for simplicity describe our method for the Bose-Hubbard model, but the same results apply to any type of interactions for systems with translational symmetry.

\emph{Universal bosonic functionals.---}
In this work we consider Hamiltonians of the form 
\begin{align}
\label{hamiltonian}
\hat H_{W}(\hat h) \equiv \hat h+\hat W,
\end{align}
 with a one-particle term $\hat h = \hat t +\hat v$, containing the kinetic energy and the external potential terms, and the two-particle interaction $\hat W$. The ground-state energy and  1RDM follow for any choice of the one-particle Hamiltonian $h$ from the minimization of the total energy functional $\mathcal{E}_h[\g]=  \mbox{Tr}[h \g] + \mathcal{F}_W[\g]$. The functional $\F_W[\gamma]$ is universal in the sense that it depends only on the fixed interparticle interaction $W$, and not on the one-particle Ha\-miltonian $\hat h$. Hence, determining the functional $\F_W[\gamma]$ would in principle entail the simultaneous solution of the universal correlation part of the ground state problem for any Hamiltonian $H_W(h)$. By  writing the ground-state energy as $E(h)\equiv \min_{\G}\mbox{Tr}_N[H_W(h)\G ]$, and using the fact that the expectation value of $h$ is determined by $\g$, one can replace the functional $\mathcal{F}_W[\gamma]$ by the well-known constrained search approach \cite{LE79}:
\begin{equation}\label{levy}
 \mathcal{F}_W[\g]  = \min_{\G\mapsto \g} \mbox{Tr}_N[W\G ]\,,
\end{equation}
where $\G\mapsto \g$ indicates that the minimization is carried out over all $\G$ whose 1RDM is $\g$. The main challenge of this approach is that the set of $\G$ such that $\G\mapsto \g$ is in general extremely complex to characterize, and so far only partial results are known for quasi-extremal, two-particle or translational invariant fermionic systems \cite{Schilling_2020,PhysRev.101.1730,Schilling2019,Giesbertz2020,PhysRevA.102.020401}. Even in the extremely popular DFT, the constrained search over many-body wave functions integrating to the same electronic density (i.e., $\Psi \rightarrow \rho$) is rarely explicitly carried out  \cite{doi:10.1021/acs.jpclett.8b02332}.

\begin{figure}[b!]
\begin{tikzpicture}
 \node (img) {\includegraphics[scale=0.1]{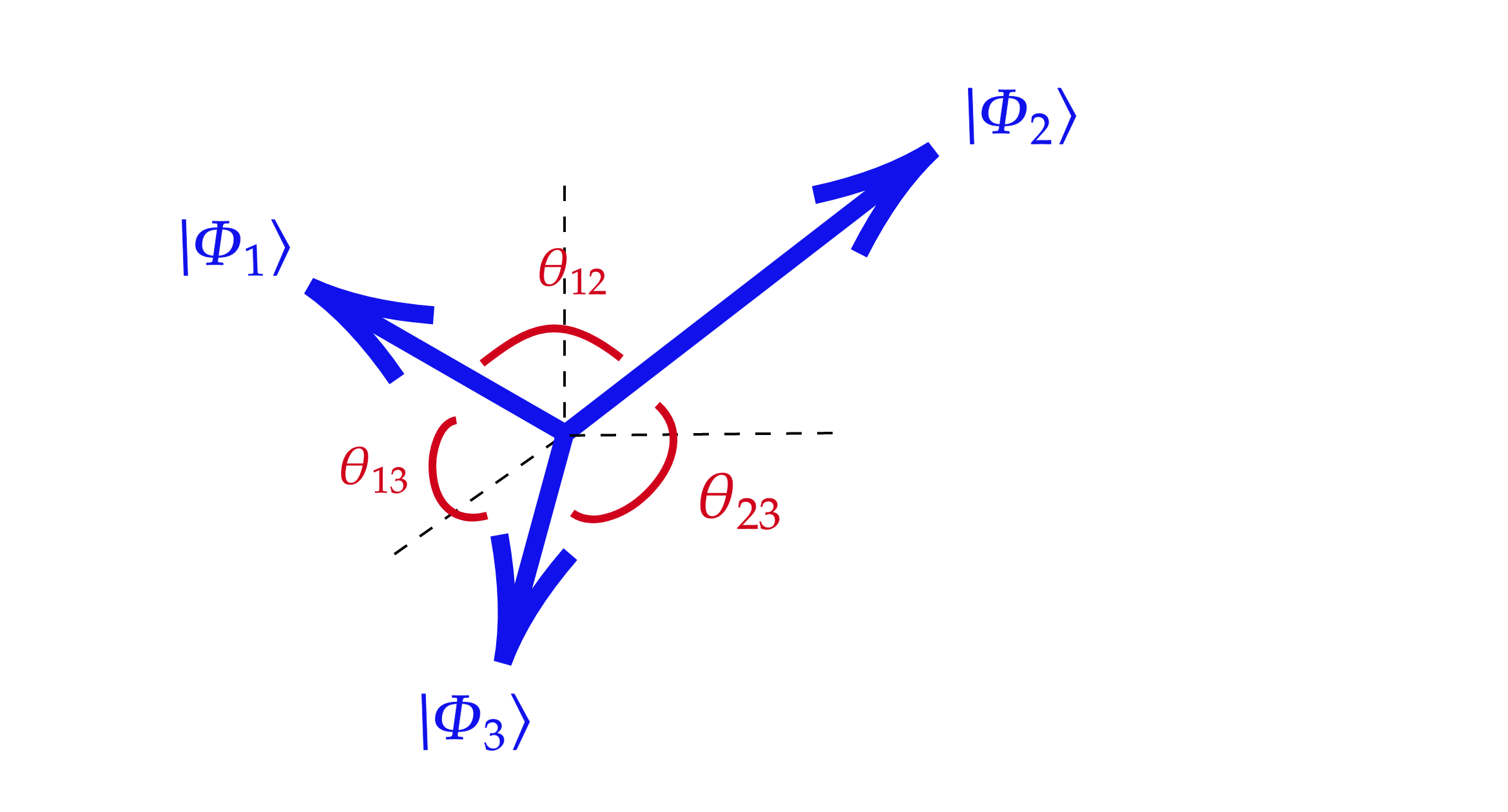}};
  \node[left=of img, node distance=0cm, anchor=center, xshift=7cm,yshift=-1.5cm,font=\color{black}] {\rotatebox{0}{\scriptsize$\gamma = \begin{pmatrix}
\bra{\Phi_1}\Phi_1\rangle & \bra{\Phi_1}\Phi_2\rangle & \bra{\Phi_1}\Phi_3\rangle\\
\bra{\Phi_2}\Phi_1\rangle & \bra{\Phi_2}\Phi_2\rangle & \bra{\Phi_2}\Phi_3\rangle\\
\bra{\Phi_3}\Phi_1\rangle & \bra{\Phi_3}\Phi_2\rangle & \bra{\Phi_3}\Phi_3\rangle\\
\end{pmatrix}$}};
 \end{tikzpicture}
  \caption{Representation of 3 wave func\-tions in the Hilbert space of $N-1$ particles giving place to a 1RDM $\gamma$. While the magnitude of the vectors is $||\Phi_i||^2 = \gamma_{ii}$, the angles they form satisfy
$\bra{\Phi_i}\Phi_j\rangle = \sqrt{\gamma_{ii}\gamma_{jj}} \cos(\theta_{ij})$.}
\label{fig1} 
\end{figure}

To shed some light on the problem let us represent $\g$, the 1RDM of a $N$-boson real wave function $\ket{\Psi}$, with respect to a set of creation and annihilation operators 
\begin{align}
\gamma_{ij} = \bra{\Psi} \hat b^\dagger_i \hat b_j \ket{\Psi}\,,
\end{align}
 and assume that the dimension of the one-particle Hilbert space is $M$. Let us also define $M$ ($N-1$)-particle wave functions $\ket{\Phi_j} \equiv \hat b_j \ket{\Psi}$,
which satisfy by definition the condition
\begin{align}
\label{condition}
\bra{\Phi_i}\Phi_j\rangle = \gamma_{ij}\,.
\end{align}
 The meaning of these non-normalized wave functions is clear: while their magnitude equals the diagonal entries of $\gamma$ (i.e., $\bra{\Phi_j}\Phi_j\rangle = \gamma_{jj}$), the angles they form correspond to the non-diagonal entries of $\gamma$. Indeed, since
$\bra{\Phi_i}\Phi_j\rangle = ||\Phi_i|| ||\Phi_j||  \cos(\theta_{ij}) = \sqrt{\gamma_{ii}\gamma_{jj}} \cos(\theta_{ij})$ we have
\begin{align}
\cos(\theta_{ij}) = \frac{\gamma_{ij}}{\sqrt{\gamma_{ii}\gamma_{jj}}} \,.
\end{align}
The bound of the non-diagonal entries: $ |\gamma_{ij}|^2 \leq \gamma_{ii}\gamma_{jj}$, is the Cauchy-Schwarz inequality for operators, and known to be a representability condition for $\gamma$ \cite{GR19}. The condition  $\sum_j \hat b_j^\dagger \ket{\Phi_j} = N \ket{\Psi}$ suggests that the minimizer of the minimization \eqref{levy} can be written as a set of $M$ vectors in the Hilbert space $\mathcal{H}_{N-1}$  of $N-1$ particles, such that their angles and magnitudes are determined by Eq.~\eqref{condition} (see  Fig.~\ref{fig1}). We now exploit this first insight to explicitly carry out the constrained search approach  and find the universal functional of the Bose-Hubbard model, a workhorse in the context of ultracold bosonic a\-toms \cite{PhysRevLett.81.3108}. 

\emph{Bose-Hubbard model.---}  The Hamiltonian of the Bo\-se-Hubbard model reads:
\begin{equation}\label{dimer}
H = -t \sum_{\langle ij \rangle}
\hat b_i^\dagger \hat b_j+\frac{U}2\!\sum^M_{j=1}\! \hat{n}_j(\hat{n}_j-1)\,,
\end{equation}
where the operator $\hat b^\dagger_{j}$ ($\hat b_{j}$) create (annihilate) a boson on site $j$, and $\hat n_{j}$ is the corresponding number ope\-ra\-tor. The first term in Eq.~\eqref{dimer} describes the hopping between two sites while the second  one is the interacting term $\hat W = \frac{U}2\sum_{j}\! \hat{n}_j(\hat{n}_j-1)$. Since the problem is determined by $N$ spinless bosons and $M$ sites, we write for the functional $\mathcal{F}_{N,M}[\gamma]$. For a given $\gamma$ 
let us take the minimizer of the functional \eqref{levy} and call it $\ket{\Psi_\gamma} \in \mathcal{H}_N$, the $N$-particle Hilbert space. Using the prescription discussed above let us define $M$ ($N-1$)-particle wave functions $\ket{\Phi_{\gamma,j}} \equiv \hat b_j \ket{\Psi_\gamma} \in \mathcal{H}_{N-1}$, which satisfy by definition the condition \eqref{condition}. Due to the translational invariance of the Bose-Hubbard Hamiltonian \eqref{dimer}, these wave functions are all normalized to the filling fac\-tor, namely, $\bra{\Phi_{\gamma,i}}\Phi_{\gamma,i}\rangle = N/M$. The functional is given by $\mathcal{F}_{N,M}[\gamma] = \sum_i \bra{\Phi_{\gamma,i}}\hat n_i \ket{\Phi_{\gamma,i}}$,
 using $\hat n_i(\hat n_i-1) = \hat b^\dagger_i \hat n_i \hat b_i$. As shown in the Appendix \ref{appb}, any rotation of the states $\ket{\Phi_{\gamma,i}}$ in the subspace spanned by themselves: $\mathcal{G}_{\gamma} = {\rm span}\{\ket{\Phi_{\gamma,1}},\dots,\ket{\Phi_{\gamma,M}}\}$, will give an energy greater  or equal than the energy $\mathcal{F}_{N,M}[\gamma]$. As a consequence, we rewrite the constrained search approach in Eq.~\eqref{levy} as  
\begin{align}
\label{univ1}
\mathcal{F}_{N,M}[\gamma] = \min_{\{\Phi_i\}\in \mathcal{G}_\gamma} \sum_i \bra{\Phi_i}\hat n_i \ket{\Phi_i}\,,
\end{align}
subject to $\bra{\Phi_i}\Phi_i\rangle = N/M$ and $\bra{\Phi_i}\Phi_j\rangle = \gamma_{ij}$. This indicates that the constraint in Eq.~\eqref{levy} can be transferred to the subspace $\mathcal{G}_{\gamma}$. As we will see below, this result leads to a quite efficient optimization problem for the functional.

\textit{Exact functional of the dimer.---} As a first illustration of this novel approach let us take the simple case of the Bose-Hubbard dimer with two particles ($N=M=2$). The states of the  Hilbert space can be written as two occupations: $\ket{n_L,n_R}$. For the 1-boson Hil\-bert space we choose as a basis: $\{\ket{1,0},\ket{0,1}\}$. We are interested in the minimum of $\bra{\Phi_L}\hat n_L \ket{\Phi_L} + \bra{\Phi_R}\hat n_R \ket{\Phi_R}$, such that $\bra{\Phi_L}\Phi_R\rangle  = \gamma_{LR} \equiv\cos(\theta)$. We write these two wave functions as $\ket{\Phi_L} = \sin(\theta_L) \ket{1,0} +  \cos(\theta_L) \ket{0,1}$ and $\ket{\Phi_R} = \cos(\theta_R) \ket{1,0} +  \sin(\theta_R) \ket{0,1}$. As a result of the corresponding minimization \eqref{univ1}, the three angles are related: $\theta_L = \theta_R = (\pi/2 - \theta)/2$, and the universal functional equals to:
\begin{align}
\label{new}
\mathcal{F}_{2,2}(\theta) = 2 \sin^2\left(\frac{\pi}4  - \frac{\theta}2 \right) = 1 -\sin(\theta)\,,
\end{align}
which is one of the few analytical results for a universal functional that can be found in the literature \cite{Cohen1,Pastor2011b}.

\emph{Machine learning.---} Despite the spectacular rise of ma\-chi\-ne learning in the study of quantum many-body systems, no implementation is known so far for the theory of reduced density matrices \cite{RevModPhys.91.045002,doi:10.1080/23746149.2020.1797528}. One of the reasons for this lack of progress is the  large amount of constraints swarming in functional theories. We now discuss how our findings will facilitate learning the universal functional of bosonic systems. Notice that a more appealing way of writing the func\-tio\-nal in Eq.~\eqref{univ1} is the following: Let us choose a basis for the vector space $\mathcal{G}_{\gamma}$, say: $\{\ket{\mathbf{m}}\in \mathcal{H}_{N-1}\}$. A set of wave functions of the sort needed in the minimization  \eqref{univ1} can now be written as $\ket{\Phi_{j}} =\sum d_{j\mathbf{m}} \ket{\mathbf{m}}$. The condition of Eq.~\eqref{condition} reads: $\mathbf{d} \mathbf{d}^\dagger =  \gamma$, where we have defined the matrix: $[\mathbf{d}]_{j\mathbf{m}} = d_{j\mathbf{m}}$. Using the singular value decomposition for such a matrix we have $\mathbf{d} = \mathbf{U} \mathbf{\Sigma} \mathbf{V}^\dagger$, 
with $\mathbf{U}$ and  $\mathbf{V}$ $M\times M$ unitary matrices. Since $\mathbf{\Sigma} \mathbf{\Sigma}^\dagger   = \mathbf{U}^\dagger \gamma \mathbf{U}$, it is now clear that the spectral decomposition of $\gamma$ equals to $\mathbf{\Sigma} \mathbf{\Sigma}^T$. As a consequence, we obtain $[\mathbf{\Sigma}]_{\alpha\alpha} = \sqrt{n_\alpha}$, where $\{n_\alpha\}$ are the eigenvalues of $\gamma$. Collecting these results we obtain that the exact universal functional \eqref{univ1} can be explicitly written in terms of the eigenvalues and the eigenvectors of $\gamma$ (contained in the matrix $\mathbf{U}$): 
\begin{align}
\label{result}
 \mathcal{F}_W[\g] = \sum_{\alpha\beta} \sqrt{n_\alpha n_\beta} \Delta_{\alpha\beta}(\mathbf{U},\mathbf{V})\,,
\end{align}
where $\Delta_{\alpha\beta}(\mathbf{U},\mathbf{V}) = \sum_{i,\mathbf{mm'}} u^*_{i\alpha} u_{i\beta} v_{\mathbf{m}\alpha}^*v_{\mathbf{m'}\beta} \bra{\mathbf{m}'}\hat n_i\ket{\mathbf{m}}$. The concrete form of the functional presented in Eq.~\eqref{result} is striking: for fermionic density-matrix-functional theory, the square root of the occupation numbers in Eq.~\eqref{result} is  known to be the optimal choice for Ans\"atze of the form $n_i^\alpha n_j^{1-\alpha}$, compatible with the integral relation between the one- and two-body reduced density matrices \cite{MULLER1984446,BB,PhysRevA.76.052517,Benavides-Riveros2018}, even for systems out of equilibirum \cite{,doi:10.1063/1.5109009}. As we can see, the only freedom in the functional \eqref{result} is the matrix $\mathbf{V}$, which is, unlike $\mathbf{U}$ and $n_\alpha$, not fixed by $\gamma$. We use this degree of freedom to introduce a standard optimization problem on a connected manifold $\mathcal{M}$: 
\begin{align}
\label{crucial}
\mathcal{F}_{N,M}[\g] = \min_{\mathbf{V}\in\mathcal{M}} \sum_{\alpha\beta} \sqrt{n_\alpha n_\beta} \Delta_{\alpha\beta}(\mathbf{U}_\gamma,\mathbf{V})\,,
\end{align}
where we have included a sub-index in $\mathbf{U}_\gamma$ to remember that such a matrix is defined by $\gamma$. Notice that the manifold $\mathcal{M}$ is essentially the set of special orthogonal matrices of dimension $M$, which generates the space $\mathcal{G}_\gamma$. Although the definition of such a space is far from trivial (and we will leave this question open for future research), it is possible to establish some elementary facts. For instance, in the strongly correlation regime $U/t\gg1$ with integer filling factor $\alpha =N/M$,  $\mathcal{G}_\gamma = {\rm span} \{b_i \ket{\alpha,...,\alpha}\}$. 

To make further progress on our problem, notice that optimization problems of the form $\min_{x \in \mathcal{M}} f(x)$ over a connected manifold $\mathcal{M}$ can be transformed into an unconstrained one of the form $\min_{y \in \mathbb{R}^n} f(\phi(y))$ by lifting the function $f$ to the current tangent space $T_x\mathcal{M}$ \cite{lezcano2019trivializations,JCM-39-207}. The map $\phi:\mathbb{R}^n \rightarrow \mathcal{M}$ is called a trivialization map \cite{lezcano2019trivializations}. By letting the minimization in Eq.~\eqref{crucial} to run over the set of special orthogonal matrices in dimension $M$, the relevant minimization space turns out to be an Euclidian space $\mathbb{R}^{M}$. As a result, finding the universal functional of RDMFT presents itself as an \textit{un\-cons\-trained} minimization problem.  This is the crucial and last finding of our work, as it finally allows us to compute the universal bosonic functional  by solving the problem directly over the set of orthonormal matrices.

Modern machine learning frameworks like pytorch \cite{pytorch} provide fast and rather efficient ways of performing optimizations on connected manifolds of the type we consider here. 
For the results we will present be\-low,  we have implemented the constrained minimization \eqref{crucial} in pytorch  with the constrained minimization toolkit GeoTorch \cite{Geotorch}. As a first step we implemented an minimization procedure where for each 1RDM the matrix $\mathbf{V}$ in Eq.~\eqref{crucial} is optimized to produce the universal functional. As a second step we trained a neural network as the universal bo\-so\-nic functional (see below).


\emph{Results.---} In Fig.~\ref{fig2a} we present the results for the Bo\-se-Hubbard model \eqref{dimer} for $M$ sites and ($\alpha M$) bosons, for $M=2,4,6$ and $\alpha = 1,2$. For this example, we have con\-si\-de\-red $\gamma$  of the form $\gamma_{ii} = \alpha$ and $\gamma_{ij} = \alpha\eta$, for $i\neq j$ with $0 \leq \eta \leq 1$ (this choice ensures the positive semidefiniteness of $\gamma$). The systems are fully condensated when $\eta = 1$ (i.e., an occupation number is macroscopically populated).  For comparison, all  functionals have been normalized to $0$  in the lower point (i.e., $\eta = 0$) and to 1 in the upper point ($\eta = 1$). The exact known results for $M=2$ 
in Eq.~\eqref{new} are verified in our calculations. Furthermore, we observe the existence of the Bose-Einstein force discovered in \cite{BR2020}, extended in \cite{liebert2020functional} and proved in \cite{macek2021absence}, namely, the divergence of the gradient $\partial_\eta \mathcal{F}_{N,M}(\eta)\rightarrow (N-N_{\rm BEC})^\zeta$, with $\zeta < 0$, when approaching to the condensation point (i.e., $\eta \rightarrow 1$).  The striking similarities of the functionals $\mathcal{F}_{M,M}[\gamma]$ and $\mathcal{F}_{2M,M}[\gamma]$ suggests the e\-xis\-tence of a universal functional independent of $\alpha$, up to appropriate normalizations.

\begin{figure}[t!]
\begin{tikzpicture}
 \node (img) {\includegraphics[scale=0.52]{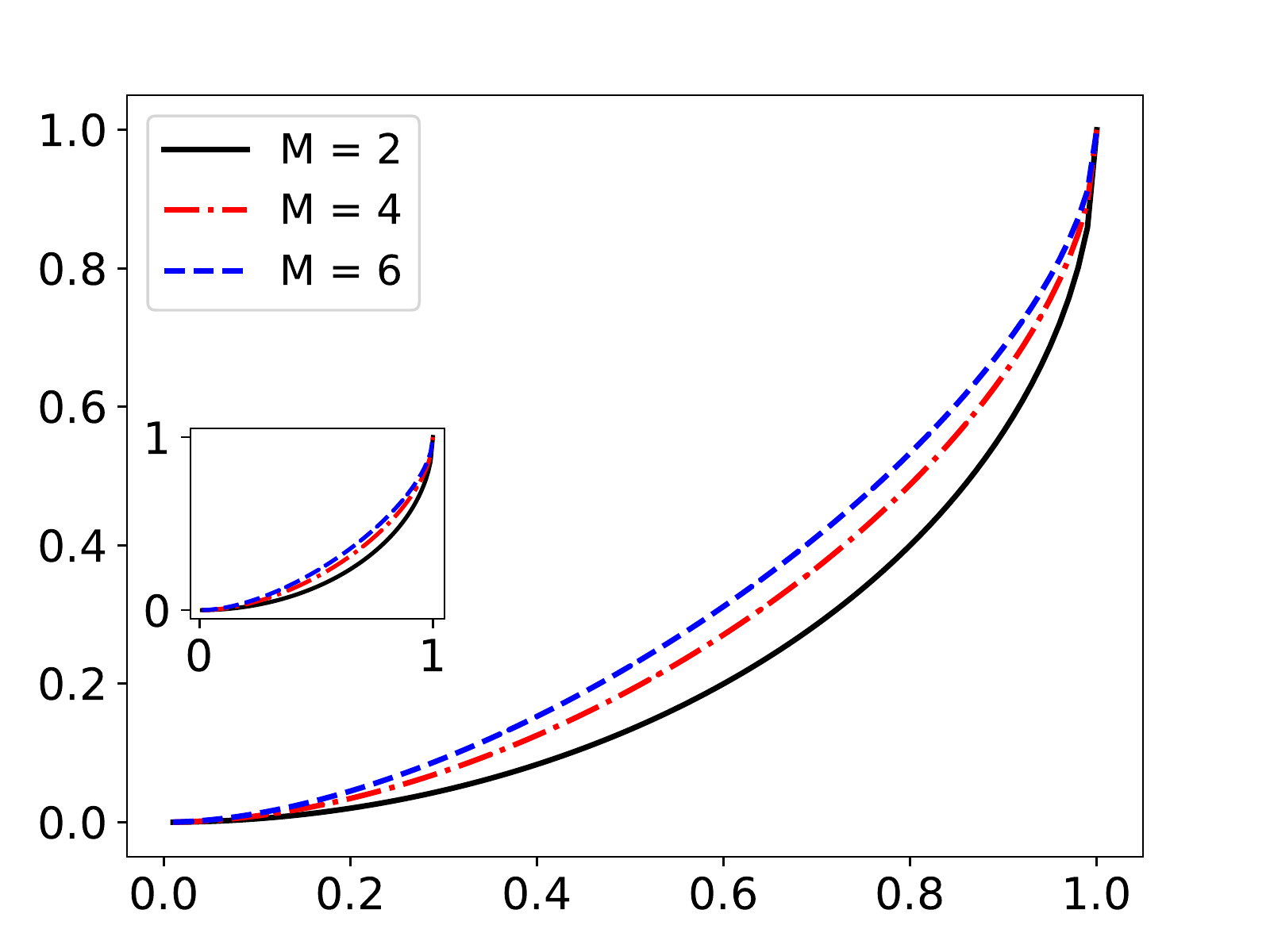}};
  \node[left=of img, node distance=0cm, anchor=center, xshift=0.9cm,yshift=1cm,font=\color{black}] {\rotatebox{90}{\large $\mathcal{F}_{\alpha M,M}(\eta)$}};
  \node[left=of img, node distance=0cm, anchor=center, xshift=5.3cm,yshift=-3.2cm,font=\color{black}] {\rotatebox{0}{\large$\eta$}};
  \node[left=of img, node distance=0cm, anchor=center, xshift=3.2cm,yshift=0.6cm,font=\color{black}] {\rotatebox{0}{\footnotesize$\alpha = 2$}};
  \node[left=of img, node distance=0cm, anchor=center, xshift=5.1cm,yshift=3cm,font=\color{black}] {\rotatebox{0}{\large$\alpha = 1$}};
 \end{tikzpicture}
  \caption{Universal functionals $\mathcal{F}_{M,M}[\gamma]$ and $\mathcal{F}_{2M,M}[\gamma]$  for the $M$-site 1D Bose-Hubbard model with filling factor $\alpha = 1, 2$, for $M= 2,4,6$ sites (see text). The functional corresponds to all 1RDM with $\gamma_{ii} = \alpha = N/M$ and $\gamma_{ij} = N\eta/M $ for $i\neq j$. All functionals are convex, as expected. For easy comparison, all functionals have been normalized to 0 in the lower point ($\eta = 0$)  and to 1 in the upper point ($\eta =1$).}
\label{fig2a} 
\end{figure}
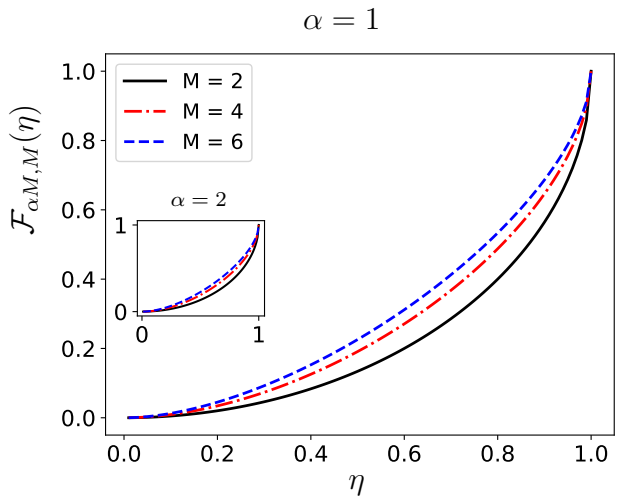

\begin{figure}[t!]
\begin{tikzpicture}
 \node (img) {\includegraphics[scale=0.5]{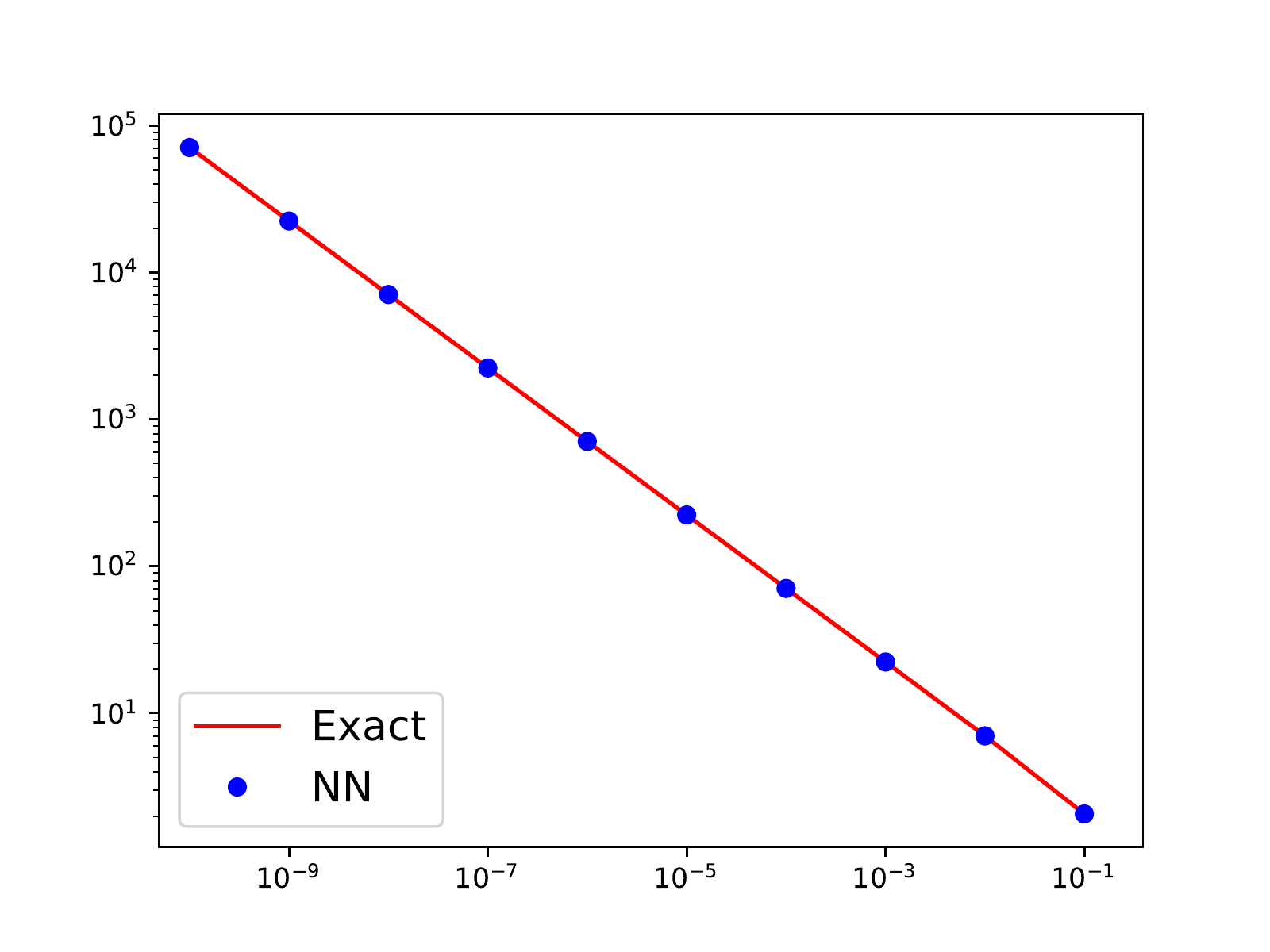}};
  \node[left=of img, node distance=0cm, anchor=center, xshift=1.2cm,yshift=0.6cm,font=\color{black}] {\rotatebox{90}{\large  $\partial_\eta \mathcal{F}(\eta)$}};
  \node[left=of img, node distance=0cm, anchor=center, xshift=5.2cm,yshift=-3.0cm,font=\color{black}] {\rotatebox{0}{\large $1-\eta$}};
 \end{tikzpicture}
  \caption{Comparison between the gradient of the functional for the  Bose-Hubbard dimer computed with the neural network and the exact analytical results as a function of $1 -\eta$ (see text). Notice the log-log scale.}
\label{fig4} 
\end{figure}

In functional theories the knowledge of the functional's form is as important as the knowledge of its derivative, as both are needed for a ground state calculation. To perform the derivative of the functional we trained a neural network to output the matrix $\mathbf{V}$ using the degrees of freedom of our 1RDM as input. This has multiple advantages. First, once the functional is trained for given particle and site numbers, it can be evaluated for any $\gamma$. Secondly, the automatic differentiation allows an exact evaluation of the gradient $\nabla_\gamma \mathcal{F}_{N,M}[\gamma]$ without further work. For the Bose-Hubbard dimer it was sufficient to use the diagonal terms $\eta = \gamma_{i(i+1)}$ and its square as inputs. The calculation was structured as follows:
\begin{equation}
\text{FCNN}_{N,M,\theta}(\eta,\eta^2,\mathbf{US}) \rightarrow \mathbf{V} \rightarrow  \mathcal{F}_{N,M,\theta}\left[\gamma\right].
\end{equation}
Here $\text{FCNN}_{N,M,\theta}$ is a fully connected network for $N$ particles and $M$ sites with the parameters $\theta$ and 
\begin{align}
\mathbf{US} = \mathbf{U}\times\begin{pmatrix}
n_0 & 0 & \cdots & 0 \\ 
0 & n_1 & \cdots & 0 \\ 
\vdots & \vdots & \ddots & \vdots \\
0 & 0 & 0 & n_M\\
\end{pmatrix}
\end{align}
calculated from the eigenvectors $\mathbf{U}$ and eigenvalues $n_i$ of $\gamma$. 
During training we minimize the functional $\mathcal{F}_{N,M,\theta}\left[\gamma\right]$ for a set of $\gamma$ in parallel. The networks used 2 hidden layers, ELU-activation functions \cite{ELU} and the output of the last layer was  used to create an orthogonal matrix through a  matrix exponential. The network for the dimer was trained on the set $\eta\in (0.005, 0.001,\dots, 0.995)$. When evaluating on the set $(0.0025, 0.0075,\dots, 0.9925)$ the maximum absolute error is smaller than $10^{-15}$. The network for $N=4$, $M=4$ was trained on the same set. Remarkably, as shown in Fig.~\ref{fig4}, for the Bose-Hubbard dimer (for which we can compare to exact results) the derivatives provided by the neural network only deviate by $\sim  10^{-13}$ from the exact results.

\begin{figure}[t!]
    \begin{tikzpicture}
  \node (img2) {\includegraphics[scale=0.45]{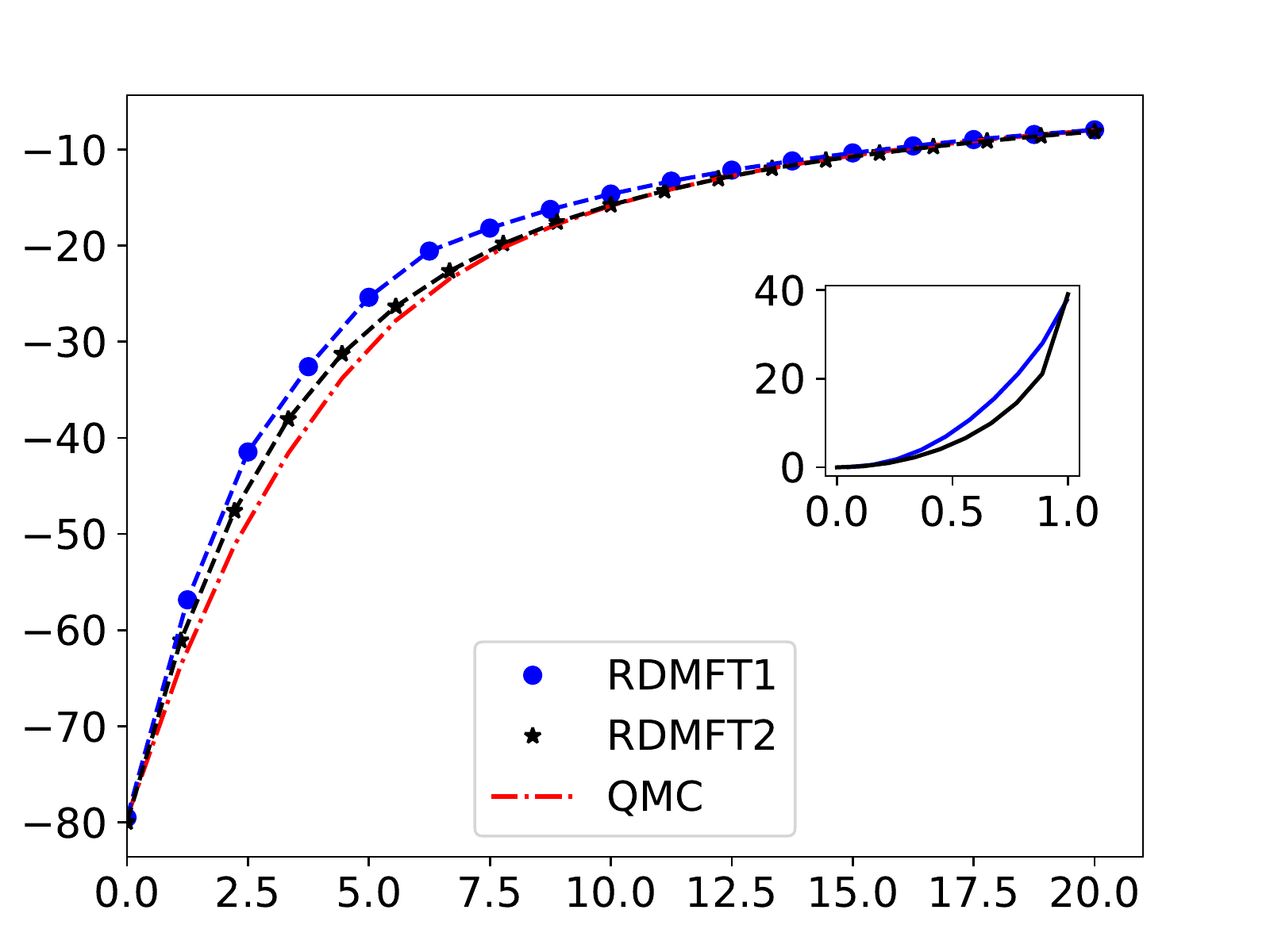}};
    \node[left=of img2, node distance=0cm, anchor=center, xshift=5.2cm,yshift=0.3cm,font=\color{black}] {\rotatebox{90}{$\mathcal{F}_{40,40}$}};
      \node[left=of img2, node distance=0cm, anchor=center, xshift=6.6cm,yshift=-0.6cm,font=\color{black}] {\rotatebox{0}{$\eta$}};
    \node[left=of img2, node distance=0cm, anchor=center, xshift=4.7cm,yshift=-2.9cm,font=\color{black}] {\rotatebox{0}{\large $U/t$}};
        \node[left=of img2, node distance=0cm, anchor=center, xshift=2.4cm,yshift=1.8cm,font=\color{black}] {\rotatebox{0}{\large \textbf{(a)}}};
 \end{tikzpicture}
\begin{tikzpicture}
 \node (img) {\includegraphics[scale=0.45]{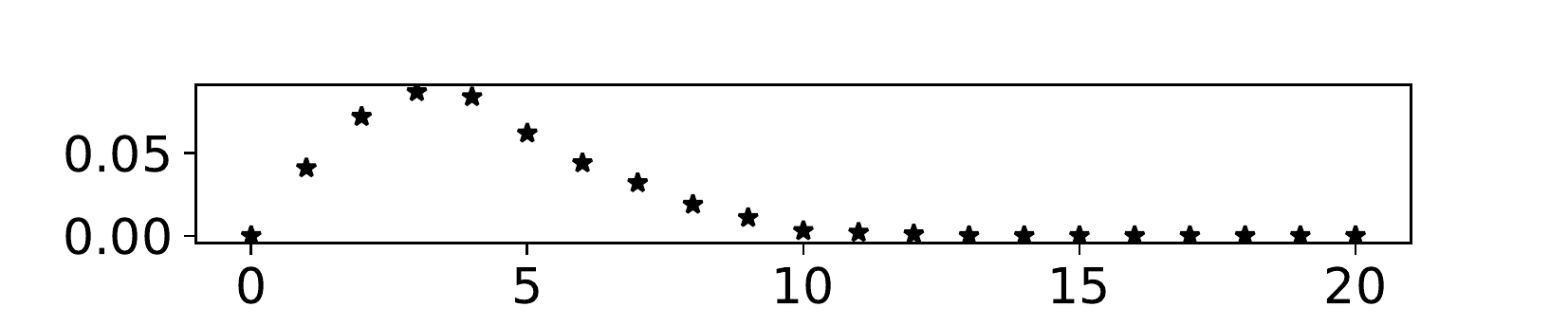}};
    \node[left=of img, node distance=0cm, anchor=center, xshift=1.0cm,yshift=0.2cm,font=\color{black}] {\rotatebox{90}{$\Delta/E_{QMC}$}};
            \node[left=of img2, node distance=0cm, anchor=center, xshift=2.4cm,yshift=0.8cm,font=\color{black}] {\rotatebox{0}{\large \textbf{(b)}}};
                \node[left=of img, node distance=0cm, anchor=center, xshift=4.7cm,yshift=-1.2cm,font=\color{black}] {\rotatebox{0}{\large $U/t$}};
     \end{tikzpicture}
  \caption{Ground-state energy of the $40$-site 1D Bose-Hubbard model with 40 bosons. In Fig.~\textbf{(a)} the energy is plotted as a function of the relative strength $U/t$. In blue the results computed using the functional \eqref{crucial} (RDMFT1) and the ansatz $\ket{\Phi_i} =  b_i \ket{1,...,1}$. In black we use as an ansatz for the functional the exact expression of the dimer appropriately rescaled (RDMFT2). In red the results computed with QMC. In the subfigure the functionals $\mathcal{F}_{N,N}(\eta)$ are plotted as a function of $\eta$, the nearest-neighbor off-diagonal value of the 1RDM. In Fig.~\textbf{(b)} the relative errors $\Delta/E_{QMC}$, where $\Delta = E_{RDMFT2}-E_{QMC}$ are plotted as a function of U/t.}
\label{fig3} 
\end{figure}
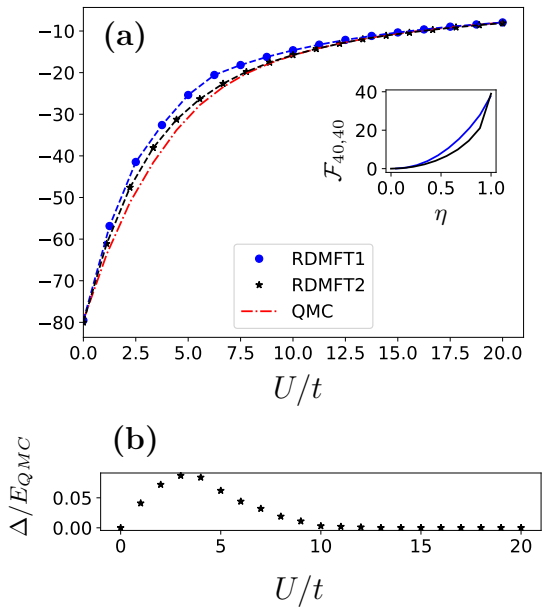

We demonstrate now that our approach allows us to compute the ground-state energy and 1RDM for a large system. Notice first that for a fixed filling factor $\alpha=N/M$, the energy of the ground state of the Bose-Hubbard model \eqref{dimer} can be computed  as the minimum of the energy functional $\mathcal{E}_{N,M}[\gamma] =  -2t\sum_i\gamma_{i(i+1)} + \mathcal{F}_{N,M}[\gamma]$. By performing the minimization $\nabla_\gamma \mathcal{E}_{N,M}[\gamma] =0$ on the domain of positive semidefinite matrices, it is then possible to compute the ground-state energy of the system. To generate the functional in that domain, we have optimized the ansatz $\gamma_{ij} = \eta^\kappa\alpha$ ($0 \leq \eta \leq 1$) with $2\leq \kappa \leq 8$, for $|j-i|>1$, with $\eta =\gamma_{i(i+1)}$. This is motivated by the fact that when $U/t\gg 0$, $\gamma_{ij} \approx 0$, $\forall i >j$, and when  $U/t\ll 1$ $\gamma_{ij} \approx \alpha$, $\forall ij$. Following this prescription we have computed the ground-state energy for the 40-site Bose-Hubbard model with 40 bosons. The dimension of the corresponding Hilbert spaces, being of the order of $5.3\times 10^{22}$, is out of reach  for exact diagonalization and prohibits performing the exact cons\-trai\-ned search approach. To solve this problem we have (i) ansatzen the space $\mathcal{G}_\gamma$, the sub\-spa\-ce generated by the kets $\ket{\Phi_i} = \hat b_i \ket{\Psi}$, by choosing $\ket{\Psi}= \ket{1,...,1}$ (RDMFT1), and (ii) used the exact functional of the dimer \eqref{new}, appropriately rescaled (RDMFT2). The energy predictions of our ma\-chi\-ne-learning functionals are quite remarkable, given the subspaces we have chosen.  Indeed, the results presented in Fig.~\ref{fig3} indicate that the predicted RDMFT results are in good agreement with the QMC energies: the errors around $U/t = 4$ are only due to the approximation of the space $\mathcal{G}_\gamma$. In order to check the quality of the approximate functionals more, we have also plotted the relative error in the last panel. We observe that this error is below 8\% and practically zero for large and weak interaction. In addition, notice that our im\-ple\-men\-tation is able to approximate the whole range of energies, not only the weakly (the sector easily described by  Bogoliubov methods) or the strongly correlation regimens. 

\emph{Conclusion.---}  In conclusion, we have demonstrated the viability of approximating universal bosonic functionals in a quite efficient way. The main ingredient of the computation is a simplification of the constrained search approach that we have introduced in this work based on the Schmidt decomposition of the wave function. This formulation of reduced density matrix functional theory (RDMFT) speeds up the design of reliable approximations for the universal functionals for systems with translational symmetry. The quality of the numerical results obtained in this work highlights the potential of RDMFT to become a competitive tool for computing properties of bosonic ground states with large dimensional Hilbert spaces. Strikingly, since RDMFT  takes into account the whole range of bosonic correlations, and does not present dimensional or sign problems, it offers a range of new possibilities. For instance, frustrated bosonic systems  can be studied in a direct manner \cite{PhysRevLett.102.017203}. Bosonic systems with impurities, composites of ultra-cold atoms, or even superconducting systems \cite{PhysRevX.10.031046,PhysRevB.99.024502,PhysRevB.99.224502,PhysRevLett.95.030405} can also potentially be addressed within this framework. As an outlook of this work, we leave open a new line of research based on extending our findings to systems with internal degrees of freedom, finite temperatures, or broken symmetries. We also expect that previous works in the context of two-body reduced density matrix \cite{PhysRevLett.83.5185,PhysRevA.69.042511} will also benefit from our approach.

\begin{acknowledgements}
We thank Jonathan Siegel, Matt Eiles, Adam Sawicki, and Ja\-kob Wolff for helpful discussions.  We are most grateful to Pe\-ter Karpov for constructive feedback and insight, and for providing us the QMC energies of the Bose-Hubbard models. M.~F.~was partially supported by the Research Fund of the University of Basel for Excellent Junior Researchers. C.~L.~B.-R.~was supported by the MPI-PKS through a next-step fellowship. 

\end{acknowledgements}

All codes to reproduce, examine and improve our proposed analysis will be made freely available online upon publication.

\appendix

\begin{widetext}

\section{The $\gamma$-representability problem}

An important problem in the theory of reduced density matrices for indistinguishable particles is the so-called $N$-representability problem, namely, which conditions should $\gamma$,  the one-body reduced density matrix (1RDM), satisfy in order to belong to at least one wave function in $\mathcal{H}_N$,  the Hilbert space of $N$ particles (fermions or bosons). To understand the problem, let us consider for a given wave function $\ket{\Psi} \in \mathcal{H}_N$  the corresponding 1RDM as:
\begin{align}
\gamma^\Psi_{ij} = \bra{\Psi}f^\dagger_i f_j \ket{\Psi}\,.
\end{align}
We have introduced a one-particle basis set $\ket{f_i} = f^\dagger_i\ket{0}$ that determines the corresponding set of creation and anhi\-li\-tation operators. The matrix $\gamma$ has the following properties: (a) it satisfies the trace condition: $\sum_i \gamma^\Psi_{ii} = N$, (b) it is hermitian, and (c) it is positive semidefinite (i.e., its eigenvalues are non-negative). For fermions the Pauli exclusion principle imposes another constraint: $\gamma \leq 1$ \cite{lieb_seiringer_2009}, and the generalized Pauli principle imposes even stronger constraints on the eigenvalues \cite{KL06,doi:10.1063/5.0031419}. Another less explored problem is the following: what is the set of wave functions giving place to the \text{same} 1RDM, namely, what is the set of wave functions
\begin{align}
\label{set}
\mathcal{S}_\gamma = \{\ket{\Psi}\in \mathcal{H}_N: \gamma_{ij} = \bra{\Psi}f^\dagger_i f_j\ket{\Psi}\}\,
\end{align}
in the Hilbert space $\mathcal{H}_N$?
The characterization of this set is of crucial importance for the ground state problem as seen from the point of view of  reduced density matrix functional theory (RDMFT) \cite{Pernal2016}. Indeed, it allows us to find the universal functional of a two-particle interaction $W$ defined as:
\begin{align}
\label{univ}
\mathcal{F}_W[\gamma] \equiv \min_{\Psi \in \mathcal{S}_\gamma} \bra{\Psi}W\ket{\Psi}\,.
\end{align}
As a consequence of this construction, the ground state of a system of indistinguishable particles driven by a Hamiltonian $ H(h)$ could be computed in a quite simple way by resorting only to the set of 1RDMs \cite{LE79}:
\begin{align}
E_0 = \min_{\Psi\in \mathcal{H}_N} \bra{\Psi}H \ket{\Psi} =
\min_\gamma \min_{\Psi \in \mathcal{S}_\gamma} \bra{\Psi}H \ket{\Psi}
= \min_\gamma  \{ \Tr[ h \gamma] + \min_{\Psi \in \mathcal{S}_\gamma} \bra{\Psi}W\ket{\Psi}\}\,,
= \min_\gamma \{ \Tr[ h \gamma] + \mathcal{F}_W[\gamma]\}\,,
\end{align}
where $ h$ contains all the 1-particle contributions to the full Hamiltonian $ H(h) =  h + W$.

\section{The Bose-Hubbard Hamiltonian}
\label{appb}

For clarity, we will work out the Bose-Hubbard model, but the results can be generalized for systems with translational symmetry. The Hamiltonian of the problem is given by:
\begin{align}
\label{ham}
 H = -t \sum_i \left(b^\dagger_i b_{i+1} + \rm{h.c.}\right) + \frac{U}2 \sum_i \hat n_i(\hat n_i -1)\,.
\end{align} 
The two-particle interaction is $W = \frac{U}2 \sum_i \hat n_i(\hat n_i -1)$ and the 1-particle Hamiltonian is $ h = -t \sum_i (b^\dagger_i b_{i+1} + {\rm h.c.})$. We fix $N$ and $M$, the number of particles and the number of sites. The filling factor is defined by $\alpha =N/M$. Notice that the Hamiltonian is translational invariant. We further choose periodic boundary conditions.

\subsection{Ground-state problem with reduced density matrices}

The ground-state energy satisfies by definition:
$E_0 =  \min_{\Psi\in \mathcal{H}_N} \bra{\Psi} H \ket{\Psi}$.
For a given wave function $\ket{\Psi}$ the expected value of the corresponding Hamiltonian reads:
\begin{align}
\bra{\Psi} H \ket{\Psi} =
 -t \sum_i  \left(\gamma^\Psi_{i(i+1)} +  c.c.\right) + \bra{\Psi}W\ket{\Psi}.
\end{align}
Yet, since the Hamiltonian \eqref{ham} is translational invariant, we have in general: $\gamma^\Psi_{ij} = \gamma^\Psi_{(i+m)(j+m)}$, for $m \geq 0$. 

As a result, the ground state problem is defined by $M/2+1$ parameters (for an even number of $M$), namely, $\gamma_{ii} = \alpha$, $\gamma_{i(i+1)} = \eta_1$, .... The 1RDM of the \textit{real} ground state can be written as a $M\times M$ matrix:
\begin{align}
\gamma(\alpha) = \begin{pmatrix}
\alpha & \eta_1 & \eta_2 &  \cdots &  \eta_3 & \eta_2  & \eta_1\\
\eta_1 & \alpha & \eta_1 &  \cdots &  \eta_4 & \eta_3  &\eta_2\\
\eta_2 & \eta_1 & \alpha &  \cdots &  * &\eta_4 & \eta_3\\
\vdots & \vdots & \vdots &  \ddots & \vdots & \vdots &\vdots\\
\eta_3 & \eta_4 & * & \cdots & \alpha & \eta_1 &\eta_2\\
\eta_2 & \eta_3 & \eta_4 &  \cdots & \eta_1 & \alpha &\eta_1 \\
\eta_1 & \eta_2 & \eta_3  & \cdots &  \eta_2 & \eta_1 &\alpha 
\end{pmatrix}\,.
\end{align}
While the entries $\gamma_{i(i+m)}=\eta_m$ for all $m > 1$ are not relevant for the computation of ground-state energy, since only nearest-neighbour hopping appears in the Hamiltonian \eqref{ham}, they are crucial for the ground-state minimization of the energy functional. Furthermore, the entries $\gamma_{ij}$ contain information about the entanglement of the modes $b_i$ and  $b_j$. Indeed, in the Mott phase $\gamma_{ij} \approx 0$ for $i\neq j$. Bose-Einstein condensation appears when  $\gamma_{ij} \approx \alpha$ for $i\neq j$. On this regard, the entries of $\gamma$ contain  relevant physical information of the problem, and impose also a challenge to the theory because the entries should be such that $\gamma$ is positive semidefinite. 

As a matter of fact, the ground-state energy can be computed by minimizing the following functional:
\begin{align}
\mathcal{E}[\gamma(\alpha)] =  -2tM\eta_1 + \mathcal{F}_W[\gamma(\alpha)]\,.
\label{eq1a}
\end{align} 
The universal functional reads as in Eq.~\eqref{univ}. Of course if we knew the expression for the functional $\mathcal{F}_W[\gamma]$ the problem would be remarkably simple: the minimum in \eqref{eq1a} could be found by simply computing the derivatives
\begin{align}
\frac{\partial}{\partial \eta_i} \big(-2tM\eta_1 + \mathcal{F}_W[\gamma(\alpha)]\big)_\alpha  = 0\,,
\label{eq2}
\end{align} 
where the subindex $\alpha$ means that it is fixed during the minimization. The minimizers $\eta_i^*$ satisfy: 
\begin{align}
\frac{\partial}{\partial \eta_1}\mathcal{F}_W[\gamma(\alpha)] = 2tM\,,
\end{align}
and $\frac{\partial}{\partial \eta_i}\mathcal{F}_W[\gamma(\alpha)] = 0$. The ground-state energy is then given by 
\begin{align}
E_0(\alpha) =  -2tM\eta_1^* + \mathcal{F}_W[\gamma^*(\alpha)]\,.
\label{eq1}
\end{align}
The main question now is if it is possible to find competitive approximations to the functional. At first sight, it seems an impossible task as it would imply the disregard of an enormous Hilbert space, whose scaling is exponential. 

We will show now that such an approach is feasible.

\subsection{Universal functional for the Bose-Hubbard model}

We tackle the problem in the following way: Let us take the minimizer of the functional \eqref{univ} and call it $\ket{\Psi_\gamma} \in \mathcal{H}_N$.  Notice first that $\ket{\Psi_{\gamma}}$ gives place to $M$ wave functions in $\mathcal{H}_{N-1}$, the Hilbert space of $N-1$ particles: 
\begin{align}
\label{newkets}
\ket{\Phi_{\gamma,i}}=b_{i}\ket{\Psi_\gamma} \,,
\end{align}
where $b_i$ is the annihilation operators of the Bose-Hubbard Hamiltonian \eqref{ham}. These wave functions satisfy $\sum_i b^\dagger_{i}\ket{\Phi_{\gamma,i}} = N \ket{\Psi_\gamma}$, $\bra{\Phi_{\gamma,i}}\Phi_{\gamma,i}\rangle = \alpha$, the diagonal of $\gamma$, and $\bra{\Phi_{\gamma,i}}\Phi_{\gamma,i+1}\rangle = \eta$. The two-particle energy of the minimizer is given by 
\begin{align}
\mathcal{F}_W[\gamma] \equiv \sum_i \bra{\Phi_{\gamma,i}}\hat n_i \ket{\Phi_{\gamma,i}}\,,
\label{partialres}
\end{align}
where we have used $\hat n_i(\hat n_i-1)$ = $b^\dagger_i \hat n_i b_i$. Now we state the following: any rotation of the states $\ket{\Phi_{\gamma,1}},...,\ket{\Phi_{\gamma,M}}$ in the space spanned by themselves  $\mathcal{G}_{\gamma} = {\rm span}\{\ket{\Phi_{\gamma,1}},\dots,\ket{\Phi_{\gamma,M}}\}$ will give an energy greater than or equal to $\bra{\Psi_\gamma}W\ket{\Psi_\gamma}$. In other words,   
\begin{align}
\sum_i \bra{\Phi'_{\gamma,i}}\hat n_i \ket{\Phi'_{\gamma,i}} \geq \sum_i \bra{\Phi_{\gamma,i}}\hat n_i \ket{\Phi_{\gamma,i}}\,,
\end{align}
for any set of states $\ket{\Phi'_1},...,\ket{\Phi'_{M}}$ that result from a rotation of the frame $\ket{\Phi_{\gamma,1}},...,\ket{\Phi_{\gamma,M}}$ in $\mathcal{G}_\gamma$. To understand the assertion, let us study in detail the Bose-Hubbard dimer with an arbitrary number $N$ of bosons. In such a case we have only two wave functions, namely, $\ket{\Phi_{\gamma,1}}$ and $\ket{\Phi_{\gamma,2}}$, with $\bra{\Phi_{\gamma,1}}\Phi_{\gamma,1}\rangle = \bra{\Phi_{\gamma,2}}\Phi_{\gamma,2}\rangle = \alpha$ and  $\bra{\Phi_{\gamma,1}}\Phi_{\gamma,2}\rangle = \alpha \eta'$. A rotation of those vectors is defined as 
\begin{align}
\ket{\Phi_1(\theta)} &= \cos(\theta) \ket{\Phi_{\gamma,1}} + \sin (\theta) \ket{\Phi^\bot_{\gamma,1}} \nonumber \\
\ket{\Phi_2(\theta)} &=  \cos (\theta) \ket{\Phi_{\gamma,2}} -\sin(\theta)
 \ket{\Phi^\bot_{\gamma,2}}\,,
\end{align}
where  $ \ket{\Phi^\bot_{\gamma,i}}$ are orthogonal wave functions to $ \ket{\Phi_{\gamma,i}}$ on $\mathcal{G}_\gamma$ defined by:
\begin{align}
    \ket{\Phi^\bot_{\gamma,1}} &= \beta \left(\ket{\Phi_{\gamma,2}}-\eta'\ket{\Phi_{\gamma,1}}\right) \nonumber \\
        \ket{\Phi^\bot_{\gamma,2}} &= \beta \left(\ket{\Phi_{\gamma,1}}-\eta'\ket{\Phi_{\gamma,2}}\right) \,
\end{align}
with $\beta = 1/ \sqrt{1-\eta'^2}$. 



Hence, taking real wave functions for simplicity, 
\begin{align}
&\bra{\Phi_1(\theta)}\hat n_1 \ket{\Phi_1(\theta)} + \bra{\Phi_2(\theta)}\hat n_2 \ket{\Phi_2(\theta)} \nonumber \\
&=   \mathcal{F}_W[\gamma]  \cos^2(\theta) + \sin^2(\theta) (\bra{\Phi^\bot_{\gamma,1}}\hat n_1 \ket{\Phi^\bot_{\gamma,1}} + \bra{\Phi^\bot_{\gamma,2}}\hat n_2 \ket{\Phi^\bot_{\gamma,2}} ) +2\cos(\theta)\sin(\theta)(\bra{\Phi_{\gamma,1}}\hat n_1 \ket{\Phi^\bot_{\gamma,1}} - \bra{\Phi_{\gamma,2}}\hat n_2 \ket{\Phi^\bot_{\gamma,2}})\nonumber \\
&= \mathcal{F}_W[\gamma] + \sin^2(\theta) (\mathcal{F}^\bot_W[\gamma]  - \mathcal{F}_W[\gamma]) +2\cos(\theta)\sin(\theta)(\bra{\Phi_{\gamma,1}}\hat n_1 \ket{\Phi^\bot_{\gamma,1}} - \bra{\Phi_{\gamma,2}}\hat n_2 \ket{\Phi^\bot_{\gamma,2}})\,,
\label{complex}
\end{align}
with $\mathcal{F}^\bot_W[\gamma] = \bra{\Phi^\bot_{\gamma,1}}\hat n_1 \ket{\Phi^\bot_{\gamma,1}} + \bra{\Phi^\bot_{\gamma,2}}\hat n_2 \ket{\Phi^\bot_{\gamma,2}}$. 
We now develop independently these terms. The second term in the last line of Eq.~\eqref{complex} can be rewritten as follows:
\begin{align}
\mathcal{F}^\bot_W[\gamma] - \mathcal{F}_W[\gamma] &=
\beta^2\left[ 
\bra{\Phi_{\gamma,2}}\hat n_1 \ket{\Phi_{\gamma,2}}
+\bra{\Phi_{\gamma,1}}\hat n_2 \ket{\Phi_{\gamma,1}}
+ \eta'^2 \mathcal{F}_W[\gamma] - 2 \eta'
\bra{\Phi_{\gamma,1}}\hat n_1 + \hat n_2\ket{\Phi_{\gamma,2}}
\right] - \mathcal{F}_W[\gamma] \nonumber \\
& =
\beta^2\left[ 
\bra{\Phi_{\gamma,2}}\hat n_1 + \hat n_2 \ket{\Phi_{\gamma,2}}
+\bra{\Phi_{\gamma,1}}\hat n_2 + \hat n_1 \ket{\Phi_{\gamma,1}}
- (1- \eta'^2) \mathcal{F}_W[\gamma] - 2 \eta'
\bra{\Phi_{\gamma,1}}\hat n_1 + \hat n_2\ket{\Phi_{\gamma,2}}
\right] - \mathcal{F}_W[\gamma] \nonumber \\
& =
\beta^2\left[ 
2 (N-1) \alpha 
- (1- \eta'^2) \mathcal{F}_W[\gamma] - 2 \eta'^2 (N-1)\alpha
\right] - \mathcal{F}_W[\gamma] = 2 (N-1) \alpha - 2 \mathcal{F}_W[\gamma]\,.
\end{align}
In the third line we have used the fact that $\ket{\Phi_{\gamma,i}}$ is an eigenfunction of the operator $\hat n_1 + \hat n_2$. The last term 
of  Eq.~\eqref{complex} can also be developed:
\begin{align}
\bra{\Phi_{\gamma,1}}\hat n_1 \ket{\Phi^\bot_{\gamma,1}} - \bra{\Phi_{\gamma,2}}\hat n_2 \ket{\Phi^\bot_{\gamma,2}}
= \beta \left[ \bra{\Phi_{\gamma,1}}\hat n_1 \ket{\Phi_{\gamma,2}} - \eta' \bra{\Phi_{\gamma,1}}\hat n_1 \ket{\Phi_{\gamma,1}}
-\bra{\Phi_{\gamma,2}}\hat n_2 \ket{\Phi_{\gamma,1}} + \eta' \bra{\Phi_{\gamma,2}}\hat n_2 \ket{\Phi_{\gamma,2}}
\right] \,,
\label{comple2}
\end{align}
which is zero, due to translational symmetry of the ground state (e.g., $\bra{\Psi_\gamma}\hat n^2_1 \ket{\Psi_\gamma} = \bra{\Psi_\gamma}\hat n^2_2 \ket{\Psi_\gamma}$). Finally, since the maximum value of the functional $\mathcal{F}_W[\gamma]$ is $N(N-1)/2$ (see supplemental material of Ref.~\cite{BR2020}) and $\alpha = N/2$, we have
that $\mathcal{F}^\bot_W[\gamma] - \mathcal{F}_W[\gamma] > 0$, 
and therefore:
\begin{align}
\bra{\Phi_1(\theta)}\hat n_1 \ket{\Phi_1(\theta)} + \bra{\Phi_2(\theta)}\hat n_2 \ket{\Phi_2(\theta)} 
> \bra{\Phi_{\gamma,1}}\hat n_1 \ket{\Phi_{\gamma,1}} + \bra{\Phi_{\gamma,2}}\hat n_2 \ket{\Phi_{\gamma,2}} \,, 
\end{align}
for $0<\theta<\pi$, which is what we wanted to prove.

A second meaningful example is the Mott phase. For $t=0$ and integer filling factor $\alpha=N/M$ the ground state is $\ket{\alpha, \alpha,...}$.
We then have $\ket{\Phi_{i}} = \sqrt{\alpha}\ket{\alpha,...,\alpha-1,...,\alpha}$ and  $\mathcal{F}_W[\gamma] = M(\alpha-1)\alpha$. A rotation of any pair of those vectors, e.g., $\ket{\Phi_i(\theta)} = \cos(\theta)\ket{\Phi_{i}}+ \sin(\theta)\ket{\Phi_{j}}$, $\ket{\Phi_j(\theta)} = \cos(\theta)\ket{\Phi_{j}}- \sin(\theta)\ket{\Phi_{i}}$ and $\ket{\Phi_k(\theta)} = \ket{\Phi_{k}}$, for $k \neq i,j$, results in the new energy ($i'= j$ and $j'= i $): 
\begin{align}
\mathcal{F}_W(\theta) \equiv \sum_i \bra{\Phi_{i}(\theta)}\hat n_i \ket{\Phi_{i}(\theta)} &= \mathcal{F}_W[\gamma] +  \sin^2(\theta)\sum_{i}
 \bra{\Phi_{i}(\theta)}\hat n_{i'} - \hat n_i\ket{\Phi_{i}(\theta)} = \mathcal{F}_W[\gamma] + 2 \sin^2(\theta) \alpha > \mathcal{F}_W[\gamma]\,,
\end{align}
for $0<\theta<\pi$. 

 This result allows us to change the \textit{constrained} search approach in \eqref{univ} by the following more appealing \textit{unconstrained} functional in the Hilbert space $\mathcal{H}_{N-1}$:
\begin{align}
\mathcal{F}_W[\gamma] &\equiv \min_{\{\Phi_i\}\in \mathcal{G}_\gamma} \sum_i \bra{\Phi_i}\hat n_i \ket{\Phi_i}\,, \nonumber \\ &  {\rm s.t.}
\quad \bra{\Phi_i}\Phi_j\rangle = \gamma_{ij}\,.
\end{align}
While there is still a representability constraint in $\mathcal{G}_\gamma$ that cannot be lifted, this construction will facilitate the design and training of a neural network as the universal functional of bosonic RDMFT. Before showing this, we employ our novel approach to explicitly compute the universal functional of the Bose-Hubbard dimer with 2 bosons, which is one of the few (or perhaps the only) analytical result for the universal bosonic functional existing in the literature.

\begin{figure}[t!]
 \includegraphics[scale=0.25]{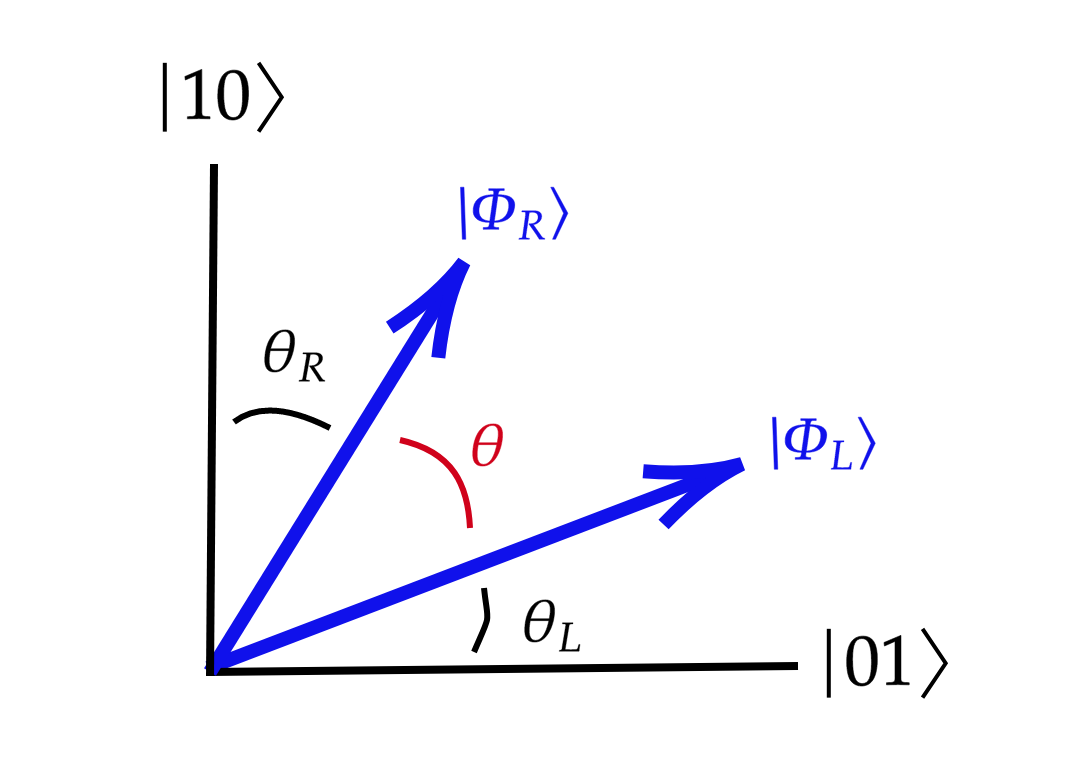}
  \caption{Schematic representation of the new unconstrained search approach introduced in this work. Here we represent the two vectors needed to find the universal functional of the Bose-Hubbard dimer with 2 bosons (see Eq.~\eqref{vecs}). The vectors $\ket{\Phi_L}$ and $\ket{\Phi_R}$ live in $\mathcal{H}_1$, the Hilbert space of 1 particle. The basis of the space is $\{\ket{01}, \ket{10}\}$. The minimum of the functional \eqref{funnew} is attained when $\theta_L = \theta_R = (\pi/2-\theta)/2$, and  the value of the functional is  found to be $\mathcal{F}(\theta) = 2 \sin^2(\pi/4 - \theta/2) = 1- \sin(\theta)$.}
\label{fig1a} 
\end{figure}

\subsection{The unconstrained search approach for the Bose-Hubbard dimer}

In this section we focus on the case $N = 2$ for the Boson-Hubbard dimer, whose Hamiltonian reads ($U >0$)
\begin{equation}
H = -t (b_L^\dagger b_R+b_R^\dagger b_L) +\frac{U}2\!\sum_{j=L/R}\! \hat{n}_j(\hat{n}_j-1),
\label{hamilt}
\end{equation}
where the operators $b^\dagger_{j}$ and $b_{j}$ create and annihilate a boson on the sites $j=L/R$, and
$\hat n_{j}$ is the corresponding particle-number ope\-ra\-tor. The dimension of the Hilbert space is 3 with basis set $\{\ket{2,0},\ket{1,1},\ket{0,2}\}$, with the configuration states defined by:
\begin{align}
\label{states}
\ket{n,N-n} = \frac1{\sqrt{n!(N-n)!}} (b^\dagger_L)^n(b^\dagger_R)^{N-n}\ket{0}.
\end{align}
The ground state belonging to such a space reads: $\ket{\Psi} = \alpha_0 \ket{2,0} + \alpha_1\ket{1,1} + \alpha_0 \ket{0,2}$. The functional to be op\-ti\-mi\-zed is:
\begin{align}
\label{fun}
\mathcal{J}[\Phi_L,\Phi_R] = \bra{\Phi_L}\hat n_L \ket{\Phi_L} + \bra{\Phi_R}\hat n_R \ket{\Phi_R}\,,
\end{align} 
such that $\ket{\Phi_i}\in\mathcal{H}_1$, $\bra{\Phi_L}\Phi_R\rangle  = \gamma_{LR}$ and $\bra{\Phi_i}\Phi_i\rangle  = 1$. We can write (see Fig.~\ref{fig1a}):
\begin{align}
\ket{\Phi_L} = \cos(\theta_L) \ket{0,1} +  \sin(\theta_L) \ket{1,0} \qquad {\rm and}  \qquad
\ket{\Phi_R} = \cos(\theta_R) \ket{1,0} +  \sin(\theta_R) \ket{0,1}.
\label{vecs}
\end{align}
Therefore $\bra{\Phi_L}\Phi_R\rangle = \cos(\theta_L) \sin(\theta_R)+ \sin(\theta_L) \cos(\theta_R)$. By defining $\gamma_{LR} = \cos(\theta)$, we have $\sin(\theta_L+\theta_R) = \cos(\theta)$. As a result, $\theta_L + \theta_R + \theta = \pi/2$, and $\sin(\theta_R) = \cos(\theta_L + \theta)$. Our functional \eqref{fun} then reads:
\begin{align}
\label{funnew}
\mathcal{J}(\theta,\theta_L) = \sin^2(\theta_L) + \cos^2(\theta + \theta_L).
\end{align}
The minimum is attached when $d\mathcal{J}(\theta,\theta_L)/d\theta_L=0$. An elementary calculation gives as a solution $\sin^2(2\theta_L^*) = \cos^2(\theta)$ for the minimizer $\theta_L^*$, which results in
\begin{align}
\mathcal{F}(\theta) = \mathcal{J}(\theta,\theta_L^*)  &= \sin^2(\theta^*_L) + \cos^2(\theta)\cos^2(\theta^*_L) + \sin^2(\theta^*_L)\sin^2(\theta) - \cos^3 (\theta)\sin(\theta) \nonumber \\ 
&=  1-\sin(\theta) = 1 - \sqrt{1 -\cos^2(\theta)} = 1 - \sqrt{1 -\gamma_{LR}^2}\, ,
\label{sol}
\end{align}
which is the result found several times in the literature for bosons and fermions \cite{Cohen1,Pastor2011b,Carrascal_2015,BR2020}. We compare the 
function in Eq.~\eqref{funnew} w.r.t.~the solution of Eq.~\eqref{sol} in Fig.~\ref{fig2a} for $\theta = 2\pi/7$. 

There is still a quicker way of computing the same result. In $\mathcal{H}_1$ the number operators can be written as $\hat n_L = \ket{10}\bra{10}$ and $\hat n_R = \ket{01}\bra{01}$. Hence, it is obvious that the minimum of \eqref{funnew} is reached when $\theta_L = \theta_R = (\pi/2 - \theta)/2$ and the value of the functional is just $2 \sin^2[(\pi/2 - \theta)/2]=1- \sin(\theta)$.

\begin{figure}[t!]
 \includegraphics[scale=0.4]{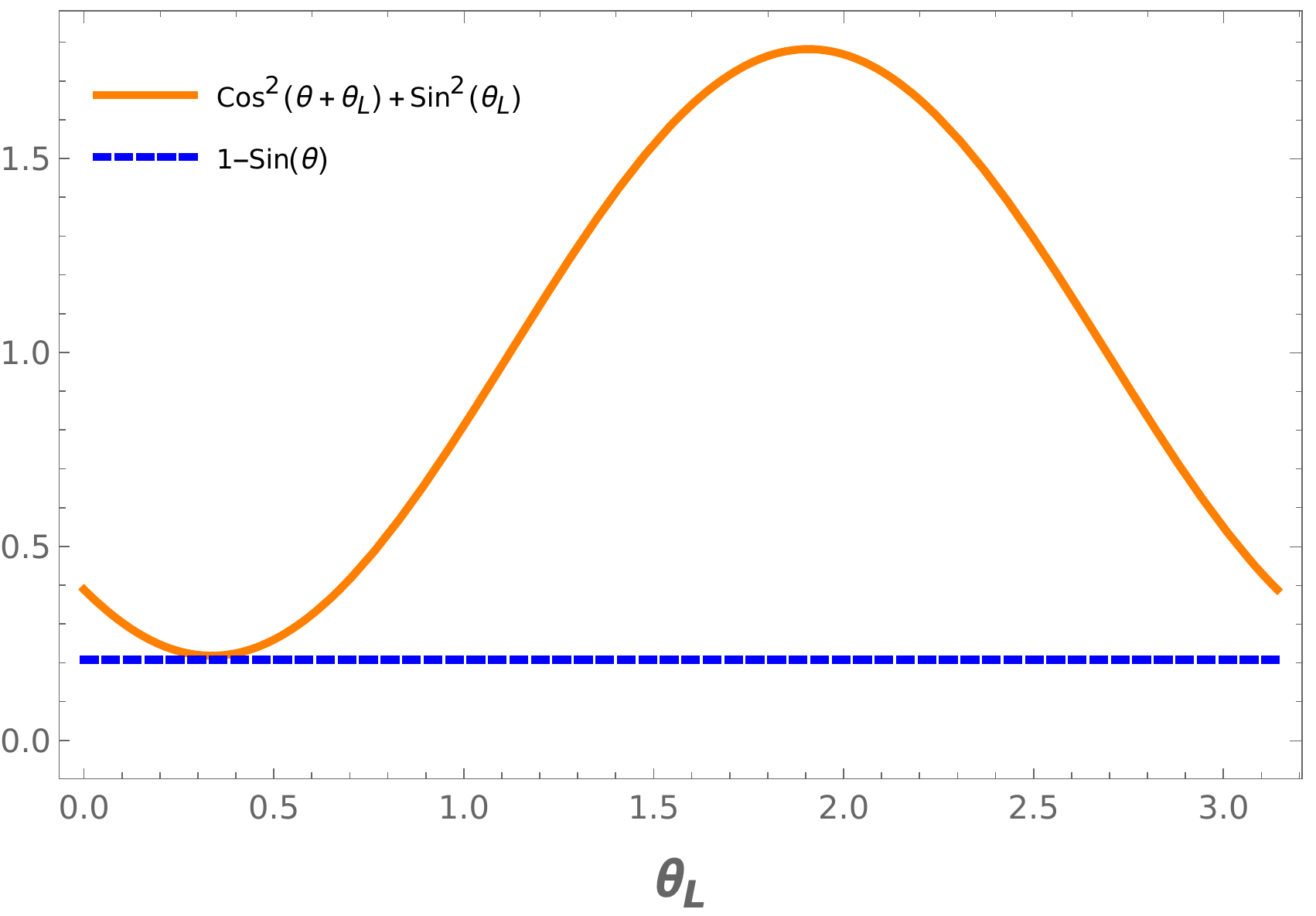}
  \caption{Plot of the unconstrained search for $\gamma_{LR}= \cos(\theta)$ with $\theta = 2\pi/7$. The minimum of the functional \eqref{funnew} is reached at $\theta_L^* = 3\pi/28$ and the value of the functional is $0.21$.}
\label{fig2a} 
\end{figure}

\subsection{The unconstrained search approach for large number of modes}

Expanded in a basis for $\mathcal{G}_{\gamma}$ the kets in Eq.~\eqref{partialres} can be written as 
\begin{align}
\label{ketsa}
\ket{\Phi'_{\gamma,i}}= \sum_{\mathbf{m}} d_{i,\mathbf{m}} \ket{\mathbf{m}}\,.
\end{align}
Then, for each $\gamma$ there is a square matrix $\mathbf{d}_{i,\mathbf{m}}$ that can be singular-value decomposed as $\mathbf{d} = \mathbf{U} \mathbf{D} \mathbf{V}^\dagger$, where $\mathbf{U}$ is an $M \times M$ complex unitary matrix, $\mathbf{D}$ is an $M \times M$ diagonal matrix with non-negative real numbers on the diagonal, and $\mathbf{V}$ is a $M \times M$  complex unitary matrix. The non-diagonal entries of the corresponding 1RDM $\gamma$ can be computed as:
\begin{align}
\gamma_{ij} = \bra{\Psi} b^\dagger_{i} b_j \ket{\Psi} = \langle \Phi_{\gamma,i} \ket{\Phi_{\gamma,j}}\,.
\end{align}
Hence, 
\begin{align}
\gamma = \mathbf{d} \mathbf{d}^\dagger =
\mathbf{U} \mathbf{D} \mathbf{V}^\dagger \mathbf{V} \mathbf{D}^T \mathbf{U}^\dagger
= \mathbf{U} \mathbf{D} \mathbf{D}^T \mathbf{U}^\dagger.
\end{align}
Therefore $\mathbf{D} \mathbf{D}^\dagger = \mathbf{U}^\dagger  \gamma \mathbf{U}$. As a consequence, $\mathbf{U}$ diagonalizes the matrix $\gamma$, and $\mathbf{D} \mathbf{D}^T$ is a diagonal matrix with the eigenvalues of $\gamma$ in the entries. Therefore the diagonal entries of $\mathbf{D}$ are $\sqrt{n_1},\sqrt{n_2},...,\sqrt{n_M}$, the square root of the eigenvalues of $\gamma$. The kets in Eq.~\eqref{ketsa} can then be explicitly written as functions of $\gamma$:
\begin{align}
\label{ketsb}
\ket{\Phi_{\gamma,i}}= \sum_{\alpha,\mathbf{m}} u_{i\alpha} \sqrt{n_\alpha} v_{\mathbf{m},\alpha}^* \ket{\mathbf{m}}\,.
\end{align}
The connection with the original wave function is striking:
\begin{align}
\ket{\Psi_\gamma} = \frac1{N} \sum_i b^\dagger_i \ket{\Phi_{\gamma,i}}= \frac1{N} \sum_i b^\dagger_i  \sum_{\alpha,\mathbf{m}} u_{i\alpha} \sqrt{n_\alpha} v_{\mathbf{m},\alpha}^* \ket{\mathbf{m}}
= \frac1{N} \sum_\alpha \sqrt{n_\alpha} \, \tilde{b}^\dagger_\alpha  \ket{v_\alpha}\,,
\label{schmidt}
\end{align}
where $\tilde{b}^\dagger_\alpha  = \sum_i b^\dagger_i  u_{i\alpha}$ and $\ket{v_\alpha} = 
\sum_\mathbf{m} v_{\mathbf{m},\alpha}^* \ket{\mathbf{m}}$. Notice that the 1RDM $\gamma$ fixes  $\tilde{b}^\dagger_\alpha$ and $\sqrt{n_\alpha}$, but not $\ket{v_\alpha}$. Quite remarkable, we can recognize in the expression \eqref{schmidt} the well-known Schmidt decomposition of a bipartite system: in this case the system of 1 and $(N-1)$ indistinguishable particles. Such a decomposition can be used to study fermionic entanglement as quantum resource \cite{PhysRevA.102.042410,PhysRevLett.120.240403} or to describe the correlated electron dynamics in strong laser fields within exact factorization of the wave function \cite{PhysRevLett.118.163202}. As its is discussed in Section \ref{NN}, the matrix  $\mathbf{V}$ is used to  engineer the bosonic functionals by training a neural network to produce $\mathbf{V}$ as a function of $\gamma$.

\section{Neural Networks}
\label{NN}

The hyperparameters used in our neural networks are the following:    


\begin{table}[!h]
 \centering      
{
\begin{tabular}{c|c| c c c  c c c c c c c}
  Hyperparameters: & N=2,M=2 & N=4, M=4\\ 
\hline
  optimizer & AdamW & AdamW \\
  momentum & 0.9& 0.9 \\
  weight-decay & 1e-06& 1e-06 \\
  learningrate &  0.00003 & 0.00001 \\
  epochs & 10000 & 20000 \\
  hidden layer size & 20, 20 & 400, 400\\
\end{tabular}}
\label{table:litioON678} 
\end{table}

\end{widetext}

\bibliography{Refs}
\end{document}